\documentclass[twoside]{article}
\usepackage{fleqn,espcrc2}

\newcommand{\be}{\begin{equation}}
\newcommand{\ee}{\end{equation}}
\newcommand{\ba}{\begin{eqnarray}}
\newcommand{\ea}{\end{eqnarray}}
\newcommand{\nn}{\nonumber}
\newcommand{\namu}{\nabla_\mu}

\newcommand{\av}[1]{{\mathrm{av}}_{#1}}
\newcommand{\Aop}{\mathcal{A}}
\newcommand{\beff}{\beta_{\mathrm{eff}}}
\newcommand{\Cav}{\mathcal{C}}
\newcommand{\Lap}{\triangle}

\newcommand{\Seff}{S_\mathrm{eff}}
\newcommand{\tr}{\mathrm{tr}\,}

\newcommand{\leftNabla}{\stackrel{\leftarrow}{\nabla}}

\newcommand{\AmS}{{\protect\the\textfont2
  A\kern-.1667em\lower.5ex\hbox{M}\kern-.125emS}}

\title{1-Loop improved lattice action for the nonlinear $\sigma$-model
\thanks{
Presented in Lattice 99 by G. Palma \newline e-mail: gpalma@lauca.usach.cl}
}
\author{M. Bartels, G. Mack\address{II. Institut f\"ur theoretische Physik,
                  Universit\"at  Hamburg \\
        D-22761 Hamburg, Luruper Chaussee 149, Germany},
and G. Palma\address{Departamento de Fisica,Universidad de Santiago de Chile,\\ 
                Casilla 307, Correo 2, Santiago, Chile}}
\begin{document}
%
%
\begin{abstract}
In this paper we show the Wilson effective action for the 2-dimensional 
$O(\mathcal{N}+1)$-symmetric lattice 
nonlinear $\sigma$-model computed in the 1-loop approximation for the nonlinear
choice of blockspin $\Phi(x)$, $\Phi(x)= \Cav\phi(x)/{|\Cav\phi(x)|}$,where
$\Cav$ is averaging of the fundamental field $\phi(z)$ over a square $x$
of side $\tilde a$.

The result for $S_{eff}$ is composed of
the classical perfect action with a renormalized coupling
constant $\beta_{eff}$, an augmented contribution from a Jacobian,
and further genuine 1-loop correction terms. Our result extends Polyakov's
calculation which had furnished those contributions to the effective action
which are of order $\ln \tilde a /a$, where $a$ is the lattice spacing of the
fundamental lattice. An analytic approximation for the background field 
which enters the classical perfect action will be presented elsewhere
\cite{BMP99}.

\vspace{1pc}
\end{abstract}
%
\maketitle
\section{Introduction}

Effective lattice actions  $S_{eff}$ in the sense of Wilson
are perfect actions in the sense that they reproduce the long 
distance behaviour  of a theory with a much larger UV cutoff. 
Different approximations have been considered before \cite{KMP96,GMXP96,H97}. 

Our computation of  $S_{eff}$ in 1-loop approximation identifies genuine 
1-loop corrections beyond the appearance of a running coupling constant in
the classical perfect action, when terms are included which are not 
$O(\ln \tilde a/a)$. 
Details and an
 analytical  approximation for $\Psi $ as a function of $\Phi $ are found 
in ref. \cite{BMP99}.
\section{Definitions}

The model lives on a square lattice $\Lambda$ of lattice spacing $a$
with points typically denoted  $z,w,\dots .$
We use lattice notations 
so that $\int_z(\dots)\equiv a^2\sum_z(\dots)\mapsto \int d^2z (...)$
in the continuum limit $a\mapsto 0$;  
$\hat{\mu}$ is the lattice vector of length $a$ in $\mu$-direction
($\mu=1,2$).
The field $\phi(z)\in S^\mathcal{N}$ is a ($\mathcal{N}+1$)-dimensional
unit vector. The action of the model is
\begin{equation}
  \label{act}
S[\phi]=
\frac{\beta}{2}\int_z[\namu\phi(z)]^2=-\frac{\beta}{2}\int_z\phi\Lap\phi .
\end{equation}
A block lattice $\tilde{\Lambda}$ of lattice spacing $\tilde{a}=s\cdot a$ is
superimposed ($s$, a positive integer).
Its points are typically denoted  $x,y,\dots .$
They are identified with squares of sidelength $\tilde{a}$ in $\Lambda$.

We define a blockspin $\Phi(x)$ which lives on the block lattice as a
function
$\Phi(x)=C\phi(x)$ of the fundamental field.
$\Phi(x)$ is also a ($\mathcal{N}$+1)-unit vector; therefore the operator
$C$
is necessarily nonlinear.
We choose
\begin{equation}
  \label{Cop}
  \Phi(x)=C\phi(x)\equiv {\Cav\phi(x)}/{|\Cav\phi(x)|} \ .
\end{equation}
The  linear operator  $\Cav$ averages over blocks,
\begin{equation}
  \label{Cav}
  \Cav\phi(x)=\av{z\in x}\phi(z)\equiv \tilde{a}^{-2}\int_{z\in x} \phi(z).
\end{equation}
The Wilson effective action is defined by
\be
  \label{weact}
e^{-S_\mathrm{eff}[\Phi]}=
        \int\,\mathcal{D}\phi\prod_x\delta(C\phi(x),\Phi(x))e^{-S[\phi]}\ ;
\ee
$\mathcal{D}\phi =\prod_z d\phi(z)$,where $d\phi $ is the uniform measure on the sphere $S^\mathcal{N}$,
and $\delta$ is the $\mathcal{N}$-dimensional $\delta$-function on the
sphere.

We consider a $\delta$-function constraint
 because computation
of expectation values of observables which depend on $\phi $ only through
the
blockspin $\Phi $ must then be identical whether computed with $S$ or
$\Seff$.
This prepares best for stringent tests of the accuracy of the result. 

Hasenfratz and Niedermayer \cite{HN94} 
showed numerically that much better locality 
properties of effective actions are obtained when a Gaussian is used 
in the definition of the effective action instead 
of a sharp $\delta$-function. Therefore we admit the substitution
\be
\delta(C\phi(x),\Phi(x)) \Rightarrow
J_0(\Cav \phi (x)) 
e^{-\frac {\beta \kappa}{2} ||\Cav^\perp[\Phi] \phi (x)||^2}
\ee
with $\Cav^\perp [\Phi ] \phi (x) = \Cav \phi (x) - \Phi(x)
 (\Phi (x) \cdot \Cav \phi (x)) = \Cav \phi^\perp (x)$ and  
$J_0$ as in eq.(\ref{jacIntro}) below. The $\delta$-function is recovered for
 $\kappa=\infty$.  
%
%
\section{Background field and classical action}

Given a blockspin configuration $\Phi$, let $\Psi=\Psi[\Phi]$ be that
field on
the fine lattice $\Lambda$ which extremizes $S$, resp. 
$S(\phi ) + \frac  1 2 \beta \kappa \sum_x |\Cav \phi^\perp (x)|^2 $
for $\kappa<\infty$  subject to the
constraints $|\Psi (z)|^2 = 1 $ and
\be
  C\Psi=\Phi\ .
\label{Psi}
\ee
$\Psi$ is called the background field.
The  classical perfect action is
\begin{equation}
  \label{classperfact}
 S_{cl}[\Phi ]= S(\Psi[\Phi]) + \frac {\beta \kappa}{2}
 \sum_x |\Cav \phi^\perp (x)|^2 \stackrel{\kappa=\infty}{\mapsto}S(\Psi).   \nn
\end{equation}
Here we wish to compute the 1-loop corrections.
It is convenient to regard the full effective action as a function of
$\Psi $. This is possible because
$\Phi $ is determined by $\Psi$ according to eq.(\ref{Psi}).

For large enough blocks, the background field $\Psi $ is smooth.

\section{The 1-loop approximation}

A perturbative calculation of the functional integral (\ref{weact}) for
the
effective action is not straightforward because the argument of the
$\delta$-function is a nonlinear function of the field.

To solve this problem, we find a parametrization 
$\phi (z) = \phi [\Psi , \zeta ](z)$
of an arbitrary field
$\phi $ on $\Lambda $ in terms of the background field $\Psi = \Psi [\Phi
] $
 and a fluctuation  field
obeying  $\zeta (z) \perp \Phi (x) \ \mbox{for } z \in x \ $ such that the
constraint becomes a linear constraint on $\zeta $, viz. $\Cav \zeta = 0$
for $\kappa=\infty$.

The background field is a smooth field. It re\-pre\-sents the low frequency
part
of $\phi $, while $\zeta $ adds the high frequency contributions.
$\zeta $ takes its values in a linear space.
 We decompose $\phi(z)$ in 
components $\perp$ and $\parallel$ to $\Phi (x),( x \ni z)$ and put
$\phi^\perp(z)= \Psi^\perp(z)+\zeta (z)$. 
Balaban \cite{Bal85} has shown how to find a suitable
parametrization for lattice gauge fields.

There is a jacobian $J$  to the transformation, and the result has the
form
\be
e^{-\Seff [\Phi ]} =   \int \prod_{z} d\zeta (z) \delta ( \Cav \zeta )
J(\Psi , \zeta )
e^{-S(\phi [\Psi , \zeta ])} \label{Seff}
\ee
with a Gaussian in place of $\delta$ if $\kappa <\infty$.

The 1-loop approximation yields the effective action  to order $\beta^0$.
It is obtained by expanding the action to second order and the Jacobian to
 zeroth order  in the fluctuation field. This approximates expression
(\ref{Seff}) by a Gaussian integral. The resulting $Tr \ log $ formula is
not particularly useful, though.

A first simplification is achieved  by exploiting the fact
that the background field $\Psi $ is smooth.
This is always true, for large enough blocks, because  a 2-dimensional 
Heisenberg ferromagnet has no domain walls \cite{MF67,DS75}. 
Because of the smoothness of $\Psi $ one can neglect
terms of higher order than second in $\nabla \Psi $. 

The exact 1-loop perfect action to this  order  is as follows.
\footnote{To save brackets, we adopt the notational convention that 
 derivatives acts only on the factor immediately following it.
We used vector notation, $\Psi^T$ is the row vector transpose to $\Psi$.
Note that $j_\mu(z)$ is a matrix.}
\ba
\Seff &=&  S_{cl} - \sum_x\ln J_0(\Cav \Psi (x)) - \frac 12 Tr \ln \Gamma_Q + \nn  \\
&&  \frac 12\int \left(
\nabla_\mu \Psi^T(z) \beff^1(z)\nabla_\mu \Psi(z) + \right. \nn \\
&& \left.
\beff^2(z) \frac  {\Phi([z])^T (-\Lap)\Psi (z)} {\cos\theta (z)}
\right) + \nn  \\
& & \Seff^{(2)}
+ \int_z \tr j_\mu(z)\nabla_\mu \Gamma_Q (z, z+ \hat \mu)  \
 , \label{result} \\
j_\mu (z) &=& \Psi(z)\nabla_\mu \Psi^T(z)  - \nabla_\mu\Psi(z)\Psi^T(z+\hat{\mu}) \nn \\
&& + \Psi \Psi^T \nabla_{\mu} \Psi \Psi^T(z+\hat \mu)   \ ,
\label{alphaDef}
\ea
where in the expression $\nabla_\mu \Gamma_Q (z, z+ \hat \mu)$ the derivative
acts only on the first argument, $[z]$ is the block containing $z$, the jacobian is
\be
J_0(\Cav \Psi (x) ) = \left(
|\Cav\Psi (x)|^2  - \frac {1}{\beta \kappa} + ...\right)^ \frac {\mathcal{N}} {2} 
\label{jacIntro}
\ee
and $S_\mathrm{eff}^{(2)} $ is a 
contribution from a renormalized 1-loop graph with 
2 vertices as follows
\ba
\Seff^{(2)} &=& -\frac 12 \int_z\int_w \tr \nn \\
&&\left(
\nabla_\mu \Gamma_Q(z,w) \leftNabla_\nu j_\nu^T(w)\Gamma_Q(w,z) j_\mu(z) \right. \nn \\
&& \left.
+ \nabla_\mu \Gamma_Q (z,w) j_\nu(w) \nabla_\nu \Gamma_Q (w,z)j_\mu(z) \right.  \nn \\
&& \left. 
+ \delta_{\mu \nu} \delta_{z,w} j_\mu(z)\Gamma_Q(z_\mu,w_\nu)j_\nu(w)
\right).
\label{ren1loop}
\ea

We used the notation $z_\mu = z+\hat \mu$.
The $\delta_{\mu \nu}\delta_{z,w} $-term subtracts the part which diverges
in the limit $a\mapsto 0$. The last term in the definition 
(\ref{alphaDef}) of $j_\mu $ 
can be dropped inside eq.(\ref{ren1loop})
because its contribution is of higher order  in $\nabla \Psi$.   

$\Gamma_Q$ is an $(\mathcal{N}+1) \times (\mathcal{N}+1)$ matrix 
propagator,  
\ba
\Gamma_Q &=& 
(-\Lap + \kappa  Q^T \Cav^\dagger \Cav  Q )^{-1}\ ; \\
  Q(z)  &=&  1 - \Psi (z)\Psi^T (z) + \Phi (x) \nn\\
&& \left( \Psi^T (z)[1+ \cos \theta (z)] - \Phi^T (x) \right) \, 
\label{Qhat0} 
\ea
with $\cos \theta (z) = \Psi (z)\cdot \Phi (x)\ , \ (x\ni z).$
Both coupling constant renormalizations $\beff^1$ and $\beff^2$ have a 
residual dependence on $\Psi $ through $ Q$,
so they fluctuate somewhat with $\Psi$; to leading order 
the dependence is through $\cos\theta$. Note that  $\beff^1$ is a 
$(\mathcal{N}+1) \times (\mathcal{N}+1)$ matrix,
 while $\beff^2 $ is a scalar. 
\ba
\beff^1(z) &=& \Gamma_Q(z,z)\ , \\ 
\beff^2(z) &=& -\tr [1-\Psi \Psi^T(z)]\Gamma_Q(z,z)\ .
\ea  
Finally, the last  term in eq.(\ref{result}) is a  lattice artifact.

If $\Phi $ is smooth enough, one may expand
$$ 
-\frac{1}{2} Tr \ln \Gamma_Q 
=  \int_x \int_z \Aop(z,x)\Cav (x,z)3[\cos \theta (z) -1] 
$$
  $\quad$\\[-5mm] 
$$
+   \int_{z,w}\int_{x,y} 
\Psi^\perp (z)\cdot \Psi^\perp (w) $$
$\quad$\\[-5mm]
$$ \quad  \qquad
 \left[-\Gamma_{KG}(z,w)\Lap \Aop (w,y)\Cav(y,z)\right.
$$ $\quad$\\[-6mm] 
$$
\quad  \qquad \qquad \left.
+ \Aop(z,x)\Cav(x,w)\Aop(w,y)\Cav(y,z)\right] 
+ \dots\ 
$$ 
and substitute $\Gamma_{Q={\bf 1}}\equiv\Gamma_{KG}{\bf 1}$ elsewhere. \\
$\Aop=\kappa \Gamma_{KG}\Cav^\ast$ has a finite limit as 
$\kappa \mapsto \infty$ \cite{GK80}.

\section{Recovery of Polyakov's result}

Polyakov determined the contributions to the effective action
 which are of order $\ln \tilde{a}/a$ \cite{Poly}. They do not
depend on the detailed form of the blockspin 
which fixes the infrared cutoff in the auxiliary theory with fields $\zeta$.
The presence of $\mathcal{M}^2=\kappa  Q^T \Cav^\dagger \Cav Q$
 in the high frequency
 propagator has the  effect of
an
 infrared cutoff \cite{GK80}. To get the result modulo details of the
choice of infrared cutoff,
 we may therefore replace $\mathcal{M}^2$ by a mass term $M^2
$, where $M= O(\tilde{a}^{-1})$. $\Gamma_Q$ then 
becomes translation invariant. One shows that
$
 \left(\Phi([z])^T \Lap\Psi (z)\right)/\cos\theta (z)= 
\Psi^T(z)\Lap \Psi(z)$ by the extremality condition for $\Psi$. 
Polyakov's result is now recovered because $Tr \ln \Gamma_Q$ becomes constant
and neither $\ln J_0$ nor the last two terms in eq.(\ref{result}) are 
 of order 
$\ln \tilde a / a$ .

\section*{\bf Acknowledgement}

This work was suppor\-ted in part by Deutsche Forchungsgemeinschaft, by
FONDECYT, Nr.\,1980608, and by DICYT, Nr.\,049631PA (Chile).
G.P. wishes to thank W. Bietenholz for useful discussions.


\begin{thebibliography}{9}


\bibitem{BMP99} M.\,Bartels, G.\,Mack, and G.\, Palma, in preparation.

\bibitem{KMP96} U.\,Kerres, G.\, Mack, G.\,Palma, Nucl. Phys. {\bf B467} (1996) 510.

\bibitem{GMXP96} M.\,Grie\ss l, G.\,Mack, Y.\, Xylander, G.\, Palma, 
Nucl. Phys. {\bf B477} (1996) 878.

\bibitem{H97} P.\,Hasenfratz, Nucl. Phys. Proc. Supp. {\bf 63} (1998) 53.
 
\bibitem{HN94} P.\,Hasenfratz, F.\,Niedermayer, Nucl. Phys. {\bf B414} (1994) 785.

\bibitem{Bal85} T.\,Balaban, Commun. Math. Phys {\bf 109} (1987) 249, 
esp. eq.(2.10)f; Commun. Math. Phys. {\bf 99} (1985) 75, Sect. E; Commun.
Math. Phys.  {\bf 102} (1985) 277, Sect. C 

\bibitem{MF67} M.\,Fischer, J. Appl. Phys. {\bf 38} (1967) 981.

\bibitem{DS75} R.\,L.\, Dobrushin, S.\,B.\,Shlosman, Commun. Math. Phys. {\bf 42} (1975) 31.

\bibitem{GK80} K.\,Gawedzki, A.\,Kupiainen, Commun. Math. Phys. {\bf 77}, 31--64 (1980).

\bibitem{Poly} A.M.\,Polyakov, Phys. Letters {\bf 59B} (1975) 79.

\end{thebibliography}
\end{document}